\documentclass[11pt]{article}

\newread\testifexists
\def\GetIfExists #1 {\immediate\openin\testifexists=#1
    \ifeof\testifexists\immediate\closein\testifexists\else
    \immediate\closein\testifexists\input #1\fi}

\usepackage{gthstyle}

\immediate\openin\testifexists=e
    \ifeof\testifexists\immediate\closein\testifexists\else
    \immediate\closein\testifexists\input e\fipsf

\def\Bbb#1{\setbox0=\hbox{$\tt #1$}  \copy0\kern-\wd0\kern .1em\copy0}

\def\bbf#1{\setbox0=\hbox{$#1$} \kern-.025em\copy0\kern-\wd0
        \kern.05em\copy0\kern-\wd0 \kern-.025em\raise.0433em\box0}

\immediate\openin\testifexists=a
    \ifeof\testifexists\immediate\closein\testifexists\else
    \immediate\closein\testifexists\input a\fimssym.def          
      \def\b{\beta}         
\def\d{\delta}      \def\D{\Delta}  
          \def\l{\lambda}     
\def\m{\mu}                     \def\vv{\varphi}
\def\n{\nu}         \def\j{\psi}    
     \def\s{\sigma}  
\def\t{\tau}        \def\th{\theta}  
              
      \def\W{\Omega}  

 \def\LL{{\cal L}} \def\OO{{\cal O}}
\def\DD{{\cal D}}
\def\pa{\partial} \def\ra{\rightarrow}

\def\dd{{\rm d\hspace{0.04em}}}     
\def\ii{{\rm i\hspace{0.04em}}}

\def\deff{\ {\buildrel{\rm def}\over{=}}\ }

\def\fract#1#2{{\textstyle{#1\over#2}}}
\def\ffract#1#2{\raise .3 em\hbox{$\scriptstyle#1$}\kern-.25em/
                \kern-.2em\lower .2 em \hbox{$\scriptstyle#2$}}

\def\half{\fract12}  
\def\quartje{\scriptstyle{1\over4}}

\def\part#1#2{{\partial#1\over\partial#2}}
\def\ex#1{e^{\textstyle#1}}

\newcommand{\be}{\begin{eqnarray}}
\renewcommand{\le}[1]{\label{#1}\end{eqnarray}}
\newcommand{\ee}{\end{eqnarray}}
\newcommand{\eqn}[1]{(\ref{#1})}

\newcommand{\fn}{\footnote}
\newcommand{\newsec}[1]{\section{#1}\setcounter{equation}{0}}
\begin{document}

\begin{titlepage}
\begin{center}
\hfill ITF-2002/41  \\ \hfill SPIN-2002/23  \\ \hfill {\tt
hep-th/0208054}\\ \vskip 20mm

{\Large{\bf ON PECULIARITIES AND PIT FALLS\\ \vskip 4mm IN PATH
INTEGRALS \fn{Opening Lecture given at the 7th International
Conference on Path Integrals, "From Quarks to Galaxies", Antwerp,
27 to 31 May 2002.}}}

\vskip 10mm

{\large\bf Gerard 't~Hooft}

\vskip 4mm Institute for Theoretical Physics \\  Utrecht
University, Leuvenlaan 4\\ 3584 CC Utrecht, the Netherlands
\medskip \\ and
\medskip \\ Spinoza Institute \\ Postbox 80.195 \\ 3508 TD
Utrecht, the Netherlands
\smallskip \\ e-mail: \tt g.thooft@phys.uu.nl \\ internet:
\tt http://www.phys.uu.nl/\~{}thooft/

\vskip 6mm

\end{center}

\vskip .2in
\begin{quotation} \noindent {\large\bf Abstract } \medskip \\
Path integrals can be rigorously defined only in low dimensional
systems where the small distance limit can be taken. Particularly
non-trivial models in more than four dimensions can only be
handled with considerable amount of speculation. In this lecture
we try to put these various aspects in perspective.
\end{quotation}

\vfill \flushleft{\today}
\end{titlepage}
\eject

\newsec{Introduction.}
Although the words ``Path Integral" and ``Functional Integral" are
usually treated as though they were synonyms, one might decide
that path integrals only refer to one-dimensional systems, whereas
functional integrals can be multi-dimensional --- after all, only
one-dimensional functions (functions only depending on time) can
be interpreted as paths. If that distinction were to be made, the
phrase ``functional integral" would be more appropriate for this
lecture, since path integrals in a one-dimensional target space
formally represent the solution of ordinary partial differential
equations, and as such they hardly present any formal
difficulty.\cite{FeynmanPath} In physics, our ``integration
variables" are often functions defined in a multi-dimensional base
space, and this is where problems of a fundamental nature arise.
We wish to integrate over entire sets of functions of several
variables, not just ``paths", which are functions of time only.

Functional integrals form the back bone of Quantum Field Theory,
which is a widely applied approach in theoretical physics. In
condensed matter physics both quantum mechanical and classical
statistical models can be addressed using functional integrals,
and in elementary particle physics the functional integral
provided essential insight not only in understanding the gauge
forces in what is now known as the Standard Model, but also in the
many recent advances in Super String Theory and its successor,
$M$-theory. All these successes made most theoreticians believe
that, if there are any mathematical difficulties in defining and
justifying the use of functional integrals, then that is something
only for philosophers or mathematicians to worry about; physicists
know what they are doing.

Extensive experimentation, both in condensed matter physics and in
elementary particle physics appears to vindicate this attitude,
but it does not justify blindness for the various complications
that may arise. Indeed, as we shall argue, there is little
experimental support for the use of functional integrals in
dimensions greater than four, and the exact definitions will be
extremely complex. Consequences of this situation for model
building are often underestimated.

In Chapter \ref{define}, we address the important question, what
\emph{is} a functional integral? In general, an integral is
defined to be the limit of an infinite sum; in turn, a functional
integral is the limit of an infinite number of integrations. How
do we know whether such a limit makes sense? More often than not,
the formal definition does not address this limiting procedure
properly.

In Chapter \ref{perturb}, it is shown that, apparently, the formal
definition makes a lot of sense when the integrand is nearly
Gaussian, while the non-Gaussian part is treated perturbatively.
The difficulty here is well-known: infinities arise that have to
be `renormalized'. Often, there are severe prices to be paid for
that.

For some special cases, renormalization works. In particular, if a
theory turns out to be `asymptotically free'. In Chapter
\ref{nonpert}, we show that then, under some quite plausible
assumptions, the functional integral allows for a nonperturbative
definition. This, however, would limit us to theories where the
dimensionality of target space is at most four. In higher
dimensions, one has to make much more drastic --- and less
reliable --- assumptions concerning the existence of equivalence
classes and the absence of a  mass gap at the lattice scale, which
would be necessary for considering the transition towards the
continuum limit.

Subsequently, two separate issues are briefly discussed: the
\emph{duality transformation}, popular nowadays in string theories
(Chapter \ref{dual}), and the \emph{Wick rotation} (Chapter
\ref{Wick}), which requires some detailed thoughts in the case of
quantum gravity. Finally, some conclusions (Chapter \ref{concl}).

\newsec{Defining Functional Integrals.\label{define}} In principle, it may
appear to be straightforward to provide a rigorous definition of
what functional integrals are supposed to mean. Consider a set of
functions \(A_i(x)\), where \(i\) is some discrete index that may
take \(N\) distinct values, and \(x\) is an element of some
multi-dimensional, continuous space-time, such as Minkowski space.
Next, consider an integrand of the form \be e^{\int\ii
\LL(x)\dd^nx}\ ,\le{fint} where \(n\) is the dimensionality of
space-time, and the Lagrange density \(\LL(x)\) is generally of
the form\fn{For the time being, we ignore fermionic degrees of
freedom; complications that are special to fermions are important,
but their role in the issues raised here is of a purely technical
rather than fundamental nature.} \be \LL(x)=-\half\Big(\pa_\m
A_i(x)\Big)^2-\half m_{(i)}^2\,A_i^2- \fract1{3!}\,g_{ijk}\,A_i
A_j A_k -\fract1{4!}\,\l_{ijk\ell}\,A_iA_jA_kA_\ell\ ,\le{lagr}
where \(g_{ijk}\) and \(\l_{ijk\ell}\) are the coupling
coefficients of the `theory', and \(m_{(i)}\) are possible mass
terms (here already diagonalized). In gauge theories, \(g_{ijk}\)
may contain one space-derivative with respect to \(x\).

We wish to integrate this integrand over the entire space of
functions \(A_i(x)\), but what does this mean? Ordinary integrals
of functions \(F(A)\) of a single variable \(A\) can be defined
very rigorously \cite{Lebesgue}: one takes the sum of the function
values over a discrete and finite set of points in \(A\) space
while multiplying the integrand with the separation distance \(\D
A\deff A-A'\) of neighboring points \(A\) and \(A'\). The limit
\be \lim_{\D A\downarrow 0} \sum_A\D A\, F(A)\ ,\le{ordint} is
defined to be the Lebesgue-integral \( \int\dd A\, F(A)\). If
\(A\) has an index \(i=1,\cdots,\, N\), we simply repeat the
procedure \(N\) times, so that we obtain the \(N\)-dimensional
integral, \(\int\dd^N \vec A\,F(\vec A)\). But what do we have to
do if \(A\) depends on a \emph{continuous} variable \(x\)? This
\(x\) takes an infinite number of values. What does it mean to
repeat the procedure \(\aleph_1\) times?

At first sight, the remedy appears to be a straightforward one:
just introduce a grid in \(x\) space as well. Once \( x\)-space
has been replaced by a grid with a finite number of points in it,
our ``functional" integral has been reduced to an integral in a
space with a finite number of dimensions, and the Riemann-Lebesgue
definition of the integral is applicable. This time, however, it
is far from straightforward to justify even the speculation that
the limit for infinitely dense grids in \(x\) space makes any
sense at all.

There is one circumstance where a reasonable procedure seems to be
possible. This is the case mentioned in the introduction, where
\(x\) is one-dimensional (typically a time coordinate). In this
case, we can \emph{order} the lattice points, and first consider
the integration over all lattice points left of some value \(t\)
in \(x\) space, while keeping \(A(t)=A\) fixed. Assume this
intermediate result to be some function \(\j(A,t)\). One then
finds \cite{FeynmanPath} that \(\j(A,t)\) obeys a partial
differential equation, in fact the Schr\"odinger equation, where
the Hamiltonian \(H\) is the one associated to the Lagrangian
\eqn{lagr}.

Note, however, that this result is not obtained unless we define
the \emph{ measure} of the integral with sufficient care while
taking the limit where the grid in \(x\)-space is made infinitely
dense.\cite{Wiener} The measure required is what became known as
the Wiener measure. Note also that \(H\) is expected to be the
quantum Hamiltonian, and that its relation to the classical
Hamiltonian generated by \(\LL\) might not be unambiguous due to
ordering problems. An example in case is the path integral for a
charged quantum particle in a magnetic field. Here, only a grid in
one time variable is needed, but it must be set up in a way that
is invariant under reversal of time, otherwise the resulting
Hamiltonian may come out to be non Hermitean due to unmatched
commutators. Thus, even the one-dimensional path integral contains
some pit falls.

In general, however, the difficulties for the one-dimensional case
can be kept well under control, and the theory of partial
differential equations allows for a sufficiently rigorous
treatment.

What, however, becomes of the Wiener measure when also \(x\) is
multi-dimensional? Here, one enters the subject of Quantum Field
Theory. What is needed is a Wiener measure for the
multi-dimensional case. This time, we cannot use the theory of
partial differential equations in finite-dimensional spaces, but,
at first sight, it appears that the generalization to a
multi-dimensional Wiener measure is again straightforward. All we
need to do is study very dense grids in \(x\) space. Take a
magnifying glass and study the `theory' at a very tiny distance
scale. In the one-dimensional case, this tiny distance scale was
represented by the tiny time step \(\D t\), turning into the
infinitesimal quantity \(\pa t\) in the partial differential
equation \(\pa\j/\pa t = -i\hat H\j\). \emph{Only if,} in the
multi-dimensional case, \emph{the small-distance limit would be
similarly well-behaved,} would we be able to define the functional
integral there. In practice, one studies the theory on a grid with
lattice length \(a\), and studies the limit \(a\downarrow 0\).

Let us focus on the limiting situation, confining ourselves only
to a tiny region \(R\) in space-time. \emph{If \(R\) is small
enough then, inside \(R\), our theory should be entirely
featureless}, or at least we should have the entire theory
completely under control, at all scales, inside the region \(R\).
It is only at scales much larger than \(R\), where we expect our
theory to exhibit interesting physical features.

It is difficult to imagine that this could mean anything else than
complete scale invariance inside the tiny region \(R\), which in
practice also implies conformal invariance. What we are talking
about is what in the one-dimensional theory would simply be the
domain where the solution to the partial differential equation
behaves linearly in the time interval \(\D t\). We postulate the
existence of small enough regions \(R\) in spacetime where such
triviality occurs.  Thus, the couplings \(g_{ijk}\) and
\(\l_{ijk\ell}\) should approach fixed points in the ultra-violet
region. In principle, there exists also the possibility of
\emph{periodic} behavior of these coupling strengths in their
dependence of the logarithm of the scale of \(R\). To be precise
then, we define the `bare' couplings to depend on the grid size
\(a\) in such a way that they either approach constant values or
values periodic in \(\log(a)\). Tiny deviations from the limiting
values at small but finite \(a\) should then lead to physically
interesting structure in the physical scale.

It then remains to be proven that the scaling behavior of these
extremely complicated integrals is as desired, and, as we shall
see, this requirement will present us with rather fundamental
problems, except when the number of dimensions is four or less.

As we shall argue later, not only is the periodic scaling
behavior a problematic option, so that we should really demand the
approach of a single fixed point, we shall even demand this fixed
point either to be at the origin: \(g_{ijk}\) and
\(\l_{ijk\ell}\ra0\), or that the fixed point values be very
small.

\newsec{Perturbation Expansion.\label{perturb}}
The case that we presently really do have under control is when
all couplings are sufficiently tiny to warrant a perturbative
approach. In the absence of all couplings, the integral \eqn{fint}
is exactly Gaussian, and its value is easy to compute exactly. In
the case when the Lagrangian  \( \LL \)  does contain
non-quadratic contributions, we extract a quadratic part and
expand the exponent of the remainder as a perturbative expansion
in terms of (powers of) \(g_{ijk}\) and \(\l_{ijk\ell}\). The
contributions of these terms in the perturbation series to an
amplitude are usually represented as Feynman diagrams.

Computation of the contribution of a given Feynman diagram
requires the evaluation of integrals in momentum space that often
appear to be divergent, and these divergences appear to become
more serious as we move to higher dimensions. The most harmful of
these divergences are the ultra-violet ones, which occur at large
values of the intermediate momenta \(k_\m\), and thus they refer
to the apparent unboundedness of the effects coming from very
tiny distance scales. This really means that the small-distance
limit referred to in the previous Section is not at all as
straightforward as one might think.

The situation is improved considerably by \emph{renormalization}.
In principle, what renormalization means is that the limit where
the grid line distance \(a\) tends to zero has to be taken in such
a way that the values of most, if not all, ``physical" parameters
of the theory, such as the coupling strengths \(g_{ijk}\) and
\(\l_{ijk\ell}\), as well as possible mass terms \(m_{(i)}\) in
Eq.~\eqn{lagr}, are carefully tuned to run either to infinity or
to zero during the limiting process. Also, physical operators such
as the field operators \(A_i(x)\), have to be renormalized. If no
further adjustments are needed for the limit to exist, the theory
is called `renormalizable'. In the '60s and early '70s, elementary
particle physicists worked out how renormalization works in
combination with the perturbation expansion, and the required
limiting procedure was identified for the complete set of models
that can be called \emph{perturbatively
renormalizable}.\cite{DIAGRAMMAR}

The conditions for a field theory to be perturbatively
renormalizable can be summarized very easily. Since we usually put
\(c=\hbar=1\), there is only one dimensionful unit required to
gauge the physical parameters, Usually, we take this to be a
length scale, say a cm or a Fermi. We now \emph{must} require that
all physical parameters in terms of which we need to do a
perturbation expansion, have a dimension of length raised to some
power that is either negative or zero. This can easily be
understood: this dimensionality assures that, at extremely tiny
distance scales, the effects of these coefficients can be ignored,
or they stay sufficiently small (in case the dimension were zero);
thus, the theory tends to a free theory in the far ultra-violet,
and the limiting process described earlier can be carried out
successfully.

The scaling dimension for the physical parameters of a theory can
be found as follows. In \(n\) space-time dimensions, the Lagrange
density function \(\LL\) of Eqs,~\eqn{fint} and \eqn{lagr} has the
dimensionality of \([\hbox{length}]^{-n}\), which will be
indicated as ``length dimension \(-n\)", or ``mass dimension
\(n\)".  From the first, kinetic term in Eq.~\eqn{lagr}, which
carries no physical constants, it follows that the fields \(A_i\)
have mass dimension \((n-2)/2\), and subsequently that the cubic
couplings \(g_{ijk}\) (if not associated with space-time
derivatives) will have mass dimension \be n-\fract32(n-2)=3-n/2\
,\le{dimg} and the quartic couplings have mass dimension \be
n-\fract42(n-2)=4-n\ .\le{diml} In four space-time dimensions
(\(n=4\)), these are the only couplings with non-negative mass
dimension. \emph{All renormalizable theories in four space-time
dimensions have at most quartic polynomials in their Lagrangians!}

The only reason why it took us several decades to work out the
technical details of the procedure needed to get into grips with
the small-distance structure of models for elementary particles,
and to renormalize them properly, was \emph{an apparent clash
between Lorentz invariance and local gauge-invariance.} A simple
grid in space-time would suspend Lorentz invariance, so that
horrendously complicated intermediate results would obscure the
proofs. Consequently, a battery of different, more symmetric
cut-off procedures were invented, each having more elegant
symmetry, but most of them either violated local gauge invariance,
or unitarity, while both local gauge invariance and unitarity are
necessary to ensure not only that our theories are stable, but
also that we can limit ourselves to interactions with the right
dimension. Thus, we had to find out how to \emph{link} various
different kinds of cut-off (or `regularization') procedures.

One of the most universally applicable cut-off procedures turns
out to be `dimensional renormalization'.\cite{dimrem} If we take
the number of space-time dimensions slightly away from the
physical value (which would be typically 3 or 4), the integrals
all come out finite\fn{More precisely: the finite and infinite
parts of the integrals can be separated from one another
unambiguously, and the infinite parts cancel completely upon
renormalization.}, while singularities that develop when the
physical number of dimensions is approached can be renormalized
separately. Using this dimensional renormalization, one sees that,
naturally, coupling parameters that are dimensionless in the
physical space-time dimensions, nevertheless need to be
renormalized, and that the `bare' constants to be used at the
cut-off scale tend to depend logarithmically on this scale.

\newsec{The non-perturbative definition.\label{nonpert}}
Although, by construction, the renormalization procedure was
always linked to the perturbation expansion, we now see that its
importance transcends perturbation theory. The point is that the
ultraviolet limit of the theory should be required to be virtually
non-interacting, which means that, just because perturbation
expansion can be applied there, the smooth limit condition of the
previous Section should be fulfilled. If the perturbative
renormalizability condition is not met, we know for sure that the
effective interaction strengths at short distances will run out of
the domain where they can be handled perturbatively.

The theory is renormalizable in the usual sense, if all physical
parameters have negative or zero length dimension, but this still
leaves two distinct possibilities at the extremely tiny distance
scales. The dimensionless bare coupling parameters depend on the
cut-off scale logarithmically. In the far-ultraviolet, these
so-called `running coupling parameters' may either run away from
the perturbative domain, or they all run to zero.

If the parameters run away from the perturbative domain, we
formally have the same situation as in non-renormalizable
theories. In practice, however, these models are still superior to
non-renormalizable theories, because the logarithmic scale
dependence is extremely slow as long as the couplings are weak.
This allows one to perform extremely accurate calculations even if
the mathematical basis is imperfect. One cannot make the
space-time grid infinitely dense, but clashes only occur at scales
of the order of \(\exp(-C/\l)\), where \(\l\) is a typical
coupling parameter, and if \(\l\) is sufficiently small, this is
completely negligible. A typical example of a theory where
accurate calculations can be done even if the mathematical basis
suffers from this difficulty is Quantum Electrodynamics. Most
researchers agree that the likely explanation for its successes is
that QED is an `effective' theory; particles and forces that
radically modify the physics at ultrashort distance scales have
simply been ignored.

\subsection{The asymptotically free case.}
\newcommand\der[2]{{\dd#1\over\dd#2}} Theories where all couplings
run to zero in the ultraviolet (``asymptotic freedom") are in a
superior shape. \cite{Politzer}\cite{GrossW}\cite{GtHrengroup}
Here, we believe that the short-distance behavior is completely
under control. Formally, the short-distance domain is described by
perturbation theory. If \(\m\) is the mass scale at tiny distances
(the distance scale \(\ell\) simply being defined as \(1/\m\)),
then the running couplings \(\l (\m)\), or, in gauge theories,
\(g^2(\m)\), obey equations such as \be
\m\der{}{\m}\l(\m)=-|\b_2|\l^2+\b_3\l^3+\d\b(\l)\ , \le{rengrp}
where \(\d\b(\l)\) stands for higher order terms in this equation.
We see that solutions look as follows: \be
{1\over\l(\m)}=|\b_2|\log\m+{\b_3\over|\b_2|}\log\l(\m) +
\OO\left(\int_0^{\l(\m)}\
{\dd\l\,\d\b(\l)\over|\b_2|\l^4}\right)+{1\over\l_{\,0}}\
,\le{running} where \(\l(\m)\) at the r.h.s.~must be eliminated
recursively, which is unambiguous as long as \(\l\) is small
enough. \(\l_{\,0}\) is some fixed integration constant. If here,
\(\d\b(\l)\) is interpreted as some higher order disturbance, then
we observe that all higher order effects that vanish as \(\l^4\)
or faster, for small \(\l\), will not seriously affect the
limiting expressions. As soon as the coefficients \(\b_2\) and
\(\b_3\) are known, Eq.~\eqn{running} can be used to \emph{define}
the running coupling parameter \(\l(\m)\), by fixing the
integration constant \(\l_{\,0}\). To give this definition,
\(\b_3\) was explicitly needed, because of the divergence of
\(\log\l\), whereas the details of the quartic corrections
\(\d\b(\l)\) are not needed to be known explicitly; they are to be
absorbed in the definition of \(\l_{\,0}\).

It is tempting to assume that \emph{all} higher order corrections
to the amplitudes, beyond those of the irreducible two-loop
diagrams, can therefore be absorbed into tiny redefinitions of the
coupling constant \(\l_{\,0}\), but this remains to be proven. To
illustrate how difficult this problem is: instanton effects are
known to be associated with a new, free constant of Nature, the
instanton angle \(\th\), whereas their amplitudes are proportional
to \(\exp(-C^{\rm\,nst}/\l)\), which tends to zero much faster
than \(\l^4\); we could have missed these effects if the above
definition were trusted blindly. Attempting to use our definition
of the functional integral in terms of the ``physical constant"
\(\l_{\,0}\) yields expressions of which the convergence could
still not be proven. This is a fundamental problem for theories
such as QCD, in spite of its phenomenal success at describing the
strong interactions among hadrons.

In spite of the absence of rigorous mathematical proofs, there
appears to be no serious difficulty in practice in the use of
theories such as QCD. Experimentally, the agreements are
impressive. We strongly suspect therefore that the definition of
what a functional integral is, in the case of an asymptotically
free quantum field theory, is an acceptable one.

A powerful argument in favor of the suspicion that the
mathematical definition of asymptotically free theories, starting
from a running coupling strength \(\l(\m)\), is unambiguous, is
the following: Imagine that two  \emph{different} theories existed
that both were described by the same scaling limit. It should be
possible to trace this difference to differences in the effective
interactions at tiny distance scales. Could we write a model
describing this effective interaction \(\D\LL(x)\) at small
distances? Unitarity and causality would demand that \(\D\LL\) has
the form of a strictly local (effective) Lagrangian. It should be
different from the defining Lagrangian of the theory. If the
defining Lagrangian contains all possible terms with the right
symmetry and dimensionality, the only possibilities left for
\(\D\LL\) is some effective Lagrangian of a higher dimensionality,
such as one containing higher derivatives or higher powers of the
fields. Interactions of this sort are usually called
\emph{marginal terms} in the interaction Lagrangian. Terms with
\emph{lower} dimensionality do not exist. Since marginal terms,
which have higher dimensionality than the canonical ones, would
diverge at tiny distances, while the amplitudes there had been
postulated to have the desired scaling behavior, we can exclude
the presence of such exotic effective interactions, ergo, the
theory must be a unique one.

\subsection{Theories with strong interactions in the UV limit.}

According to perturbation theory, models that are not
asymptotically free can still be described by coupling parameters
that run as a function of the scale \(\m\): \be
\m\der{}{\m}\l(\m)=\b(\l(\m))\ ,\le{beta} where the function
\(\b(\l)\) can be computed perturbatively. If \(\b\) is negative,
the theory is asymptotically free. Let us now assume \(\b(\l)\) to
be positive. After having computed the first few terms of the
expansion \(\b(\l)\) for small \(\l\), one may find that the
function appears to sport a zero at some finite value of \(\l\):
\be\b(\l_{\,0})=0\,.\le{zero} This zero then is an attractive one
in the UV direction. If the coupling has the value \(\l_{\,0}\),
the theory is scale-invariant (which usually implies also
conformal invariance). As we desire to give a rigorous definition
of the theory, we put it on a lattice, as before, and give the
`bare' coupling parameter the value \(\l_{\,0}\) (with possibly a
small correction). It is tempting then to assume that, indeed, the
amplitudes will reproduce the scale invariance. But now, there are
several problems.

First, just because the coupling is never very small, the
artifacts due to the finite lattice size are complicated to
compute, and they will be non-negligible. They may invalidate the
argument that the theory scales, so that the limit where the
lattice length \(a\) tends to zero is not well under control.

Secondly, if the coupling is strong, its actual value depends
strongly on many details in its definition. In perturbation
theory, we see this as a dependence on the subtraction procedure,
at every order of the calculations; in the lattice theory we see
this as a dependence on details of the definition of the lattice.

However, theories that have a lattice with strong interactions in
the far ultraviolet domain (that is, a very dense lattice), and
which do tend to become scale-invariant in the far infrared
(i.e.~at large distances) do exist. We know that perturbative
schemes exist where the coupling constant \(\l_{\,0}\) can be made
small but non-zero. If such a theory is put on a lattice, then we
have an example of a non-asymptotically free theory that is
\emph{presumably} well-defined.

Can such non-asymptotically free theories be unique? Again, we can
attempt to argue that any non-uniqueness would be described by
effective interactions with anomalous dimensions. This time,
however, we cannot use perturbation theory at the ultraviolet as a
guide. If the theory is `nearly' asymptotically free, that is,
\(\l_{\,0}\) is small, the argument appears to be as reliable as
in the asymptotically free case, but if  \(\l_{\,0}\) is large, we
have no clue.

Even worse will the situation be if there is no fixed point at
all, but instead a \emph{limit cycle}. This may happen in the case
of multiple coupling parameters, which in the ultraviolet domain
tend to a periodic solution of Eq.~\eqn{beta}. Here, much less is
known. If such theories do exist, and if they can be demonstrated
to be unique, then we should be able to list them as
\emph{universality classes}, much like what is done in statistical
mechanics in three space dimensions. Again, we must assume that
marginal terms can be excluded.

An important case is quantum field theory in more than four
space-time dimensions. Here, we see that all quartic terms in the
lagrangian have negative mass dimension (see Eq.~\eqn{diml}). We
must have quartic terms (if not higher) if we want the Hamiltonian
of the theory to be bounded from below, so that the system is
stable. This implies that the function \(\b(\l)\) starts off with
a term \((n-4)\l\), which is positive. To develop a zero in this
function, we need the higher order terms and hence the coupling
can never be small. This also holds if we want a limit cycle.
Thus, in more than four space-time dimensions it is inevitable to
have strong coupling at the cut-off point. One may even question
whether a zero in \(\b(\l)\) can ever occur. In any case, the UV
limit cannot be treated using perturbation theory. It can only be
treated by speculation. The only alternative would be a numerical
experiment using computers, but now the problem is that, precisely
in a large number of space-time dimensions, numerical algorithms
tend to become prohibitively slow.

A superior approach to the questions at hand is to start from some
generic lattice theory that exhibits the required symmetry
properties of the continuum theory one wants to study. Then we
should ask: is there any set of values for the various coupling
parameters such that there is a non-trivial far infrared region,
where the theory becomes scale-invariant yet non-trivial? If such
a set is found, one can subsequently consider a slight deviation
from these values, which will break the infrared scale invariance,
thus producing effective mass terms, and with those, more
non-trivial structure. Looking upon our problem this way, it is
evident that success depends on the existence of such a set of
coupling parameters. It definitely does \emph{not} exist within
the perturbative domain, if \(n>4\). One might conclude for these
reasons that the existence of any consistent quantum field theory
at \(n>4\) should be dubious. This should even include
supersymmetric theories. Supersymmetry is difficult
--- if not impossible --- to reproduce on a lattice.

But we could also speculate on the existence of fine-tuned
theories that do survive in some non-trivial manner. These will be
theories with strong interactions throughout, often with
scale-invariance and possibly with other special symmetries such
as supersymmetry. They may form universality classes. In the next
Section, we speculate that the number of distinct universality
classes could be smaller than what is suggested by writing down
Lagrangians.

\newsec{duality\label{dual}}

Again, first consider theories with a rigorous lattice cut-off. In
such systems, different systems may exhibit dual relationships.
The simplest example of such a relationship is the 2-dimensional
Ising Model, where the coupling parameter \(\b\) can be given any
value from \(-\infty\) to \(+\infty\). By rearrangement of the
primary degrees of freedom, one finds that all properties of the
model wih given \(\b\) can be mapped onto the features of the same
model, living on the dual of the original lattice\fn{The dual
lattice is obtained from the original lattice by interchanging the
plaquettes with the lattice sites.}, at the coupling \(\b^*\),
where \(\b^*\) is related to \(\b\) as \cite{KramersWannier} \be
\ex{2\b^*}=\coth\b\ .\le{ising} The relation is dual in the sense
that, when applied twice, it returns to the original model.

Relations of this sort are more general. In three dimensional
lattice theories, one finds a similar dual relation linking two
different models: the Ising Model is dually related to the \({\Bbb
Z}_2\) theory on the dual lattice. In four dimensions, the \({\Bbb
Z}_2\) theory is self-dual.

Duality on the lattice is not restricted to \({\Bbb Z}_2\)
theories, but, for its rigorous definition, it does require an
\emph{Abelian} structure. In fact, Eq.~\eqn{ising} is the simplest
example of a Fourier transform in parameter space. If the link
variable can be written as a commuting quantity \(\s\), and the
action is written as \be\sum_{\rm links\ \ell}\b(\s_\ell)\ ,
\le{abel} then the dual theory obtains the same action, but with
\(\b^*(\s'_\ell)\), defined by \be Z\,\ex{\,\b^*(\s')}=
\sum_{\s}\ex{\,i\s\,\s'}\ \ex{\,\b(\s)}\,,\le{gendual} where \(Z\)
is a normalization factor. The duality transformation
\eqn{gendual} can be generalized even more by having a
higher-dimensional \(\s\) field.

These observations would allow us to perform dual transformations
on a variety of theories, which however all have to be Abelian.
The Fourier transformation in Eq.~\eqn{gendual} is a linear
transformation, and there seems to exist no direct generalization
towards non-Abelian systems. Thus, in four space-time dimensions,
only Abelian gauge theories can be dually transformed to other
Abelian gauge theories. What we \emph{can} do for non-Abelian
theories is first integrate away their non-Abelian parts. This is
indeed exactly what is done in the procedure called Abelian
projection in QCD.\cite{AbelProj} The Abelian, or Cartesian
subgroup of the gauge group can be seen to correspond to an
ordinary Abelian gauge theory, to which the non-Abelian sector
adds not only electrically charged objects, but also magnetic
monopoles. The dual transformation then interchanges the monopoles
and the electric charges.

Unfortunately, it appears to be impossible to carry out such a
transformation procedure exactly. It is therefore quite remarkable
that nevertheless dual transformations among supersymmetric gauge
theories appear to be possible, provided that one restricts
oneself to the far infrared domain. a possible explanation for the
remarkable facts that were discovered is that the dual
transformations only hold for the universality classes, not for
the individual theories with any finite cut-off.

\newsec{The Wick Rotation.\label{Wick}}
An important calculational tool in many quantum field theories is
the Wick rotation. First, one notices that, in Feynman diagrams,
rotational symmetry can be exploited more fully by substituting
\be k_0\quad\ra\quad k_4\ =\ {\rm i}k_0\,.\le{krot} Here, this is
not more than a simple contour rotation in the complex plane of
the integration variable \(k_0\). In coordinate space, the
equivalent rotation is \be t\quad\ra\quad \t\ =\ {\rm i}
t\,.\le{trot} The functional integral expression, for instance in
a gauge theory, is then replaced as follows: \be\int\DD
A\,e^{-{\rm i}\int \quartje\,F_{\m\n}F_{\m\n}\,\dd^{n-1}\vec
x\,\dd t}\ra \int\DD A\,
e^{-\quartje\,F_{\m\n}F_{\m\n}\,\dd^{n-1}\vec
x\,\dd\t}\,.\le{functrot} In all conventional quantum field
theories, the complex integrand turns into a Gaussian integrand,
and, being the exponential of a negative quantity, the integrals
converge optimally.

In the case of the gravitational field, however, things work out
differently. At first sight, one is tempted to proceed in exactly
the same way. The substitution \(t\ra\t={\rm i}t\) can be
performed in the Einstein-Hilbert action. In the functional
integral for what should become a quantum theory of gravity, one
[erforms the switch \be \int\DD g_{\m\n}\,e^{{\rm
i}\int\sqrt{-g}\,R\,\dd^{n-1}\vec x\,\dd t}\ra\int\DD
g_{\m\n}\,e^{\int\sqrt{g}\,R\,\dd^{n-1}\vec
x\,\dd\t}\,.\le{gravrot}

The difficulty here is that the Einstein-Hilbert action, \(\sqrt
g\,R\), is not at all bounded from below, not even after rotating
all i's away. Consequently, the resulting integration expression,
Eq.~\eqn{gravrot}, makes no sense at all.

Elaborate proposals have been formulated to turn this meaningless
expression into something one can calculate; the problem is
sometimes believed to cure itself, somehow. This, however, is not
the correct answer. The correct answer is found by returning to
the roots of the procedure: rotating integration contours in the
complex plane. In ordinary perturbation theory, one sees most
easily what happens. The rule here is: if you have an integral
that converges because the integrand becomes rapidly oscillating
at infinity, you can obtain an equivalent expression that
converges faster by rotating the integration contour in the
complex plane. The rotation \emph{must} be performed in such a way
that, while rotating the contour over a variable angle \(\vv\),
the integrand converges at infinity. In practice, this means that
integration variables may be chosen to rotate in the complex
plane, but this must always be done in such a way that the
integral becomes an exponentially \emph{convergent} one.

Let us take four dimensional space-time, to be explicit. If we
rotate the metric field variables, \be\int\DD
g_{\m\n}=\prod_x\int_{C1}\dd g_{00}(x)\int_{C2}\dd g_{01}(x)\cdots
\int_{C10}\dd g_{33}(x)\,\nonumber\ee we must ensure that the
resulting integrals converge. Now, because of local gauge
invariance, we must impose a gauge condition, and add the usual
ghost term. Just as in Maxwell theory, after fixing the gauge, not
all surviving degrees of freedom are truly dynamical. Some of them
act as Lagrange multipliers. In Maxwell theory, of the four vector
components of the vector potential field \(A_\m\), one disappears
as a consequence of the gauge condition, and an other one turns
into a Lagrange multiplier to produce the Coulomb potential. Two
physical photonic degrees of freedom survive. In gravity, there
are 10 components of \(g_{\m\n}\), and 4 gauge conditions (fixing
the 4 coordinates). Of the 6 surviving fields, 4 act as Lagrange
multipilers, so that two graviton degrees of freedom remain.

In perturbation theory, one sees most clearly that some of these
Lagrange multipliers should not rotate so that the metric would
become nonnegative. Although the details depend on how the gauge
was fixed, it appears in general not to be possible to avoid
contours to rotate incorrectly, unless we keep the metric complex.
This is easy to understand: this happens \emph{because} the
Einstein-Hilbert action is unbounded!

In many gauge choices, one finds that only the \emph{conformal
factor} in the metric needs to be complex. Thus, we write
\be\int\DD g_{\m\n}\ra\prod_x\oint\dd\W\prod_{\m\n}\dd\hat
g_{\m\n}\,,\le{conf} where \be g_{\m\n}\equiv\W\hat g_{\m\n}\
,\quad\hbox{and}\quad R(\hat g_{\m\n})\equiv 0\,.\le{gtransf}

\newsec{Conclusion.\label{concl}}
The possibility to define functional integrals in more than one
space-time dimensions depends on the existence of universality
classes. In less than four dimensions, these classes are
relatively easy to define, since the far ultraviolet
(i.e.~small-distance) domain of the theory is controlled by
perturbation theory, which we know how to handle. In four
dimensions, this is also the case if we have asymptotic freedom,
or possibly if the coupling strength tends to a zero of the \(\b\)
function where it is itself small, so that one may still rely on
perturbation theory to find a useful theory.

However, in particle theory, and notably in string theory and in
supergravity, one often speculates on theories in much more than
four space-time dimensions. The possibility to \emph{define} what
a functional integral is, depends on features one can only
speculate about.

What we know for certain about our physical world is that
functional integrals successfully describe statistical systems in
three space-dimensions and elementary particles in four space-time
dimensions.

If there exist more dimensions describing physics at ultra-tiny
scales, then well-defined functional integrals in more than four
dimensions would be needed there. Such theories can only stretch
over a large domain of scales if, at the largest distance scales,
the effective interactions are either extremely weak (since the
only allowed effective interactions are marginal ones), or
extremely fine-tuned: the physical interaction parameters are very
strong, and they are tuned in such a way that the theory scales.
Models based on continuum physics but nevertheless exhibiting
interesting non-trivial interactions, in more than four
dimensions, are therefore physically unrealistic, and this may
explain why, as yet, no experimental evidence has been found in
favor of Kaluza-Klein theories for elementary particles.

The Wick rotation in quantum gravity is less enigmatic than what
is often claimed, but the real physical significance of quantum
wave functions on complex conformal factors may have to be
investigated further.

\end{document}